\documentclass[pdflatex,sn-mathphys-ay]{sn-jnl}% Math and Physical Sciences Author Year Reference Style

\usepackage{graphicx}%
\usepackage{multirow}%
\usepackage{amsmath,amssymb,amsfonts}%
\usepackage{amsthm}%
\usepackage{mathrsfs}%
\usepackage[title]{appendix}%
\usepackage{xcolor}%
\usepackage{textcomp}%
\usepackage{manyfoot}%
\usepackage{booktabs}%
\usepackage{algorithm}%
\usepackage{algorithmicx}%
\usepackage{algpseudocode}%
\usepackage{listings}%
\begin{document}

\title[Article Title]{Seasonal forecasting using the GenCast probabilistic machine learning model}

\author*[1]{\fnm{Bobby} \sur{Antonio}}\email{bobby.antonio@physics.ox.ac.uk}
\author[1,2]{\fnm{Kristian} \sur{Strommen}}
\author[1]{\fnm{Hannah M.} \sur{Christensen}}
\affil[1]{\orgdiv{Atmospheric, Oceanic and Planetary Physics}, \orgname{University of Oxford}, \orgaddress{\street{Sherrington Road}, \city{Oxford}, \postcode{OX1 3PU}, \country{United Kingdom}}}
\affil[2]{ \orgname{European Centre for Medium-Range Weather Forecasts}, \orgaddress{\street{Shinfield Rd}, \city{Reading}, \postcode{RG2 9AX}, \country{United Kingdom}}}

% 150 to 250 words
\abstract{
Machine-learnt weather prediction (MLWP) models are now well established as being competitive with conventional numerical weather prediction (NWP) models in the medium range. However, there is still much uncertainty as to how this performance extends to longer timescales, where interactions with slower components of the earth system become important. We take GenCast, a state-of-the-art probabilistic MLWP model, and apply it to the task of seasonal forecasting with prescribed sea surface temperature (SST), by providing anomalies persisted over climatology (GenCast-Persisted) or forcing with observations (GenCast-Forced). The forecasts are compared to the European Centre for Medium-Range Weather Forecasts seasonal forecasting system, SEAS5. Our results indicate that, despite being trained at short timescales, GenCast-Persisted produces much of the correct precipitation patterns in response to El Ni\~{n}o and La Ni\~{n}a events, with several erroneous patterns in GenCast-Persisted corrected with GenCast-Forced. The uncertainty in precipitation response, as represented by the ensemble, compares favourably to SEAS5. Whilst SEAS5 achieves superior skill in the tropics for 2-metre temperature and mean sea level pressure (MSLP), GenCast-Persisted achieves significantly higher skill in some areas in higher latitudes, including mountainous areas, with notable improvements for MSLP in particular; this is reflected in a higher correlation with the observed NAO index. Reliability diagrams indicate that GenCast-Persisted is overconfident compared to SEAS5, whilst GenCast-Forced produces well-calibrated seasonal 2-metre temperature predictions. These results provide an indication of the potential of MLWP models similar to GenCast for the `full' seasonal forecasting problem, where the atmospheric model is coupled to ocean, land and cryosphere models.}

% \keywords{Seasonal forecasting, Machine Learning, Weather forecasting}

\maketitle

\section{Introduction}

Recent years have seen a proliferation of machine-learnt weather prediction (MLWP) models that are competitive with conventional physics-based models at medium-range weather forecasting, for both deterministic \citep{lam_learning_2023, bi_accurate_2023, allen_end--end_2025} and probabilistic forecasts \citep{price_probabilistic_2025, lang_aifs_2024}. So far, these models have focused mainly on short- to medium-range weather forecasts. A natural question to ask is whether these models could be used to forecast to longer horizons, specifically out to seasonal timescales.

Since current MLWP models do not have an interactive ocean, seasonal forecast experiments using these models must currently prescribe an evolving ocean state as a boundary condition. In this context, a seasonal forecast experiment primarily tests the ability of the MLWP atmosphere to respond correctly to the changing ocean state. There are several benefits of such an experiment. Firstly there is the practical aspect of evaluating how skilful these models are at seasonal timescales; if successful, then MLWP forecasts would offer a means to efficiently generate large seasonal forecast ensembles and potentially produce more skilful and reliable forecasts. Rolling out to longer timescales also serves as a useful test of the physical realism of models, and how well they generalise to tasks they are not trained on. Seasonal forecasting in particular is a test of how well the MLWP model has learned to respond to other Earth system components such as the ocean. Such applications to different tasks can build trust in the output of these models, and provide insight into how general purpose the models can be. It can also reveal undesirable behaviours of MLWP models that are not apparent at shorter timescales. For example, these experiments allow an assessment of how stable the models are at long timescales, which is important since several are known to become unstable and produce unrealistic values outside of the 14-day horizon (e.g.~\cite{karlbauer_advancing_2024}). Finally, it is of direct scientific interest to understand to what extent accurately simulating short timescale weather phenomenon automatically allows longer timescale variability to be accurately simulated as well; such understanding has direct implications for, e.g., the `seamless prediction' framework~\citep{palmer2008toward, christensen2019reliable}, wherein one tries to use information about short-term weather forecasts to constrain climate projections in models.

In order to perform forecasts beyond the medium range, there are two main approaches to consider. In the `direct' approach, a machine learning model is trained to directly predict the forecast variables at the lead times of interest. There are several studies that apply this approach to subseasonal to seasonal (S2S) forecasting (up to around 6 weeks lead time, \citet{Delaunay2022,nguyen_climax_2023, liu_cirt_2025}) and seasonal forecasting \citep{pinheiro_interpretable_2025}. As an alternative to forecasting the atmosphere, others have demonstrated how machine learning models can predict key drivers of seasonal variability such as the El Ni\~{n}o/Southern Oscillation (ENSO) index \citep{ham_deep_2019, parthipan_regularization_2025}.

Alternatively, we can adopt an `autoregressive' approach, by which we mean a model that makes predictions at a daily or sub-daily level, and is rolled out to seasonal timescales. In this approach it is hoped that a MLWP model trained at relatively short timescales will learn the correct physical interactions in order to create the correct behaviour at longer timescales. There are several studies applying this approach for S2S timescales \citep{chen_machine_2024, li_tianquan-climate_2025, chen_machine_2024, weyn_ensemble_2024, ling_fengwu-w2s_2024, zhou_machine_2025}. However, to our knowledge, this autoregressive approach has been tested on seasonal timescales in only two works: \citet{kent_skilful_2025} use the ACE2 model \citep{watt-meyer_ace2_2025} to perform seasonal forecasts, with a model that is forced with SST and sea ice cover anomalies, where ensembles are created using a lagged ensemble approach. \citet{zhang_seasonal_2025} perform seasonal hindcasts with NeuralGCM \citep{kochkov_neural_2024}, similarly using persisted SST and sea ice anomalies, with a focus on forecasting tropical cyclone activity, and creating ensembles using initial condition perturbations. We note that both ACE2 and NeuralGCM were designed with climate applications in mind.

In this work, we are are interested in further exploring the autoregressive approach applied to seasonal forecasting, since it provides an interesting test of the kind of physical relationships that MLWP models can learn having being trained at short timescales. It is also a useful precursor to assess how different models could extend to climate timescales. We use GenCast \citep{price_probabilistic_2025}, a recently developed probabilistic model that achieves state-of-the-art skill in the medium range, and explore how well it performs at the task of seasonal forecasting over a four month period with prescribed sea surface temperatures. Our setup mirrors that of \cite{kent_skilful_2025} and \cite{zhang_seasonal_2025} in that persisted SST anomalies are used as boundary condition, although we also consider a forced setup where ERA5 SSTs are provided, in order to assess where forecast skill is limited by factors beyond the ocean representation. Aside from being the first application of GenCast to forecasting beyond the medium-range, our work complements existing studies in several ways. Firstly, GenCast is a probabilistic model, which in theory can learn to directly predict the correct conditional probability distribution. We may therefore expect it to produce a more reliable ensemble compared to initial condition or lagged ensembles. GenCast was also designed specifically for the medium-range, unlike NeuralGCM and ACE2. Evaluating the model on seasonal timescales may reveal biases in GenCast that are not apparent at short lead times, and evaluates whether a model designed purely for the medium-range can possibly generalise to longer timescales. Finally, given the relatively small number of studies for seasonal prediction using autoregressive models, it is a useful additional case study, to explore any potential benefits or disadvantages of using a different model.

\section{Methods}
\subsection{Machine learning model}\label{sec:mlmodel}

GenCast makes predictions at 12hr time steps, for 6 surface variables, and 6 variables at 13 pressure levels. We use the $1^{\circ}$ model since the GPU available to us was not large enough to fit the $0.25^{\circ}$ version. GenCast receives no inputs related to the land surface (e.g.~soil moisture) and, unlike ACE2 and NeuralGCM, does not take information about sea ice as an input. Each GenCast forecast is initialised on the 1\textsuperscript{st} November, and rolled out until the end of the following February. We initialise the forecasts on years 2004-2024; this is to incorporate as much data as possible that is completely unseen by GenCast (2019-2024), as well as incorporating years with a range of different conditions. Note that, even though the years 2004-2018 are within the training period for GenCast, by rolling the forecast out autoregressively to seasonal timescales, we are still exposing the model to inputs it has not seen before. These years can therefore also be considered out-of-sample for this experiment.

GenCast is run with two different ocean boundary conditions. The first setup, GenCast-Persisted, persists the ERA5 anomalies at 1\textsuperscript{st} November on top of the SST climatology for the duration of the forecast, similarly to the approach in \cite{kent_skilful_2025} and \cite{zhang_seasonal_2025}, based on the approach in \cite{zhao_retrospective_2010}. This setup is closest to a forecast setup, where the real sea surface temperatures are not known in advance. The climatology used for this experiment is the daily SST climatology calculated over the ERA5 data from 1\textsuperscript{st} January 1979 - 12\textsuperscript{th} December 2018. The second setup, GenCast-Forced, uses ERA5 sea surface temperature as input to GenCast. This serves as a useful indicator of where skill or reliability might be improved by a more accurate representation of the ocean.

\subsection{Data}\label{sec:data}
The ERA5 reanalysis dataset \citep{hersbach_era5_2020} is used as the 'ground truth', since this is the dataset GenCast is originally trained on. As a baseline forecast, we use the European Centre for Medium Range Weather Forecasts (ECMWF) SEAS5 forecasts \citep{johnson_seas5_2019}. For both SEAS5 and the GenCast experiments we use 20 ensemble members. Data is aggregated to give an average value for the boreal winter (December-February), and is detrended when calculating the anomaly correlation coefficient and reliability diagrams to remove the climate change signal. Subregions are chosen to explore the precipitation distribution response to El Ni\~{n}o / La Ni\~{n}a in Sec.~\ref{sec:precip}, taken from the regions in \cite{davey_probability_2014} for which there is a wetting or drying signal over December-February for both types of events. The subregions (using the same naming conventions as in \cite{davey_probability_2014}) are Indonesia ([10\textsuperscript{o}S-5\textsuperscript{o}N, 100\textsuperscript{o}-130\textsuperscript{o}E]), SSAfrica ([28\textsuperscript{o}-18\textsuperscript{o}S, 18\textsuperscript{o}-33\textsuperscript{o}E]) and MexUSA ([30\textsuperscript{o}-35\textsuperscript{o}N, 120\textsuperscript{o}-90\textsuperscript{o}W]). Averages are taken over land points only, with the exception of Indonesia which includes land and sea points.

\subsection{NAO index}
We calculate the North Atlantic oscillation (NAO) index as the difference in mean sea level pressure for a region around the Azores ([28-20\textsuperscript{o}W, 36-40\textsuperscript{o} N]) and around Iceland ([25-16\textsuperscript{o} W, 63-70\textsuperscript{o} N]), following \cite{dunstone_skilful_2016}. The NAO series for each forecast is centred by subtracting the mean NAO value for that series over the 20-year period. Each series is then normalised by dividing by the standard deviation of the NAO index calculated on ERA5 data. 

\subsection{Tests of significance}
\label{sec:sig}
In order to test where anomaly correlations $r_{a}$ are significantly greater than 0, we use a one-sided t-test of the test statistic $r_{a} ( n-2)^{1/2} / (1-r_{a}^2)^{1/2}$ \citep{storch_statistical_1999}, where $n$ is the number of years used to calculate the result. In order to test where there is a significant difference between the anomaly correlation of two different forecasts, we follow the approach outlined in \cite{siegert_detecting_2017}, which accounts for the fact that the two forecasts are themselves correlated. All significance results are reported at the 95\% confidence level.

\subsection{Reliability diagrams}

To calculate the reliability diagrams, the forecasts and observations are separately detrended in order to remove any climate change signal. Terciles are then calculated for each grid cell individually, allowing a calculation of probabilities for each grid cell separately.

\section{Results}

\subsection{El Ni\~{n}o and La Ni\~{n}a case studies }
\label{sec:precip}
\begin{figure}[!ht]
    \centering
    \includegraphics[width=\textwidth]{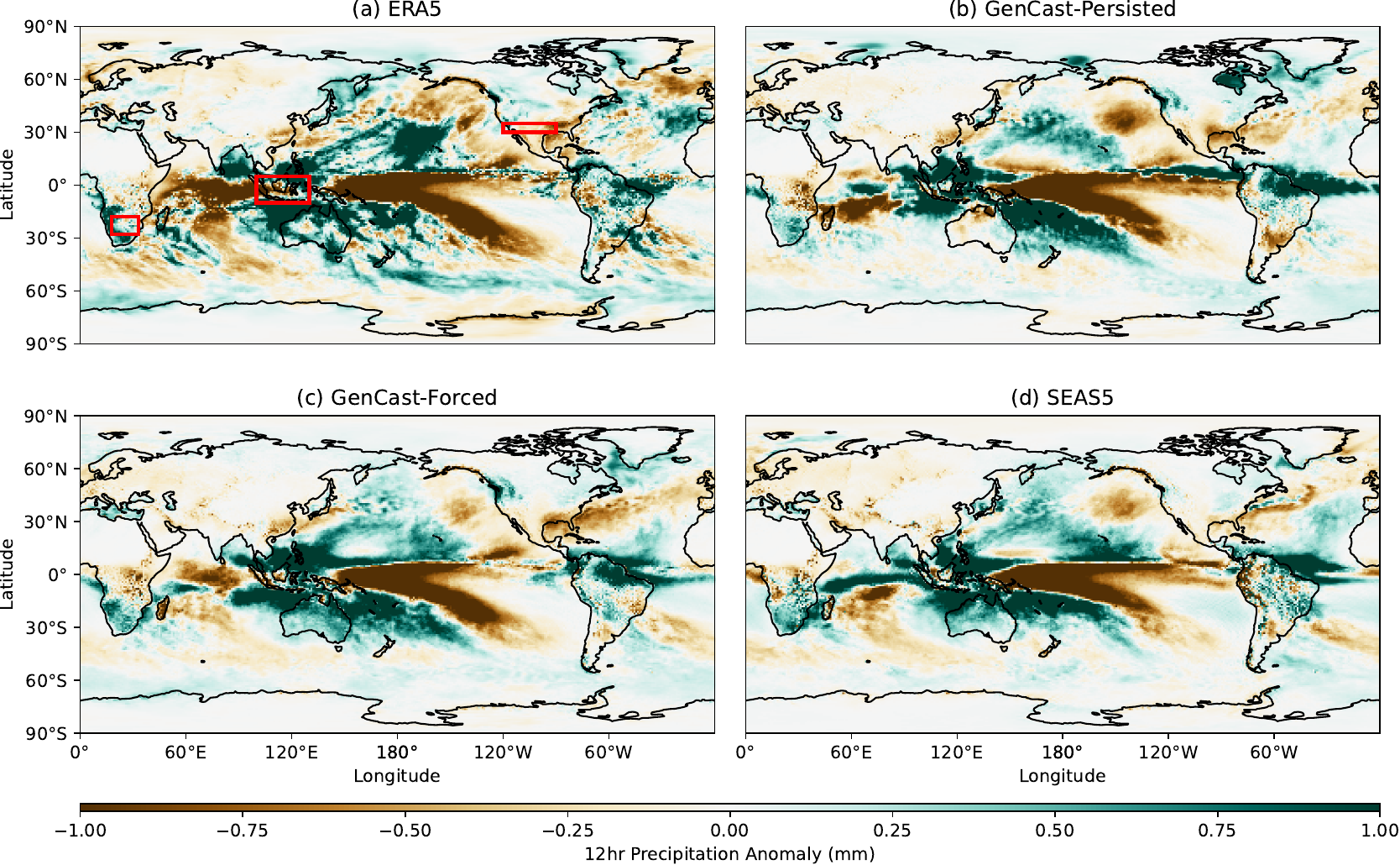}
    \caption{Seasonal 12hr precipitation anomalies for December 2010 - February 2011 (Strong La Ni\~{n}a). Red boxes indicate the subregions used in Fig.~\ref{fig:tp_distribution}. }
    \label{fig:tp_anomalies_lanina}
\end{figure}

\begin{figure}[!ht]
    \includegraphics[width=\textwidth]{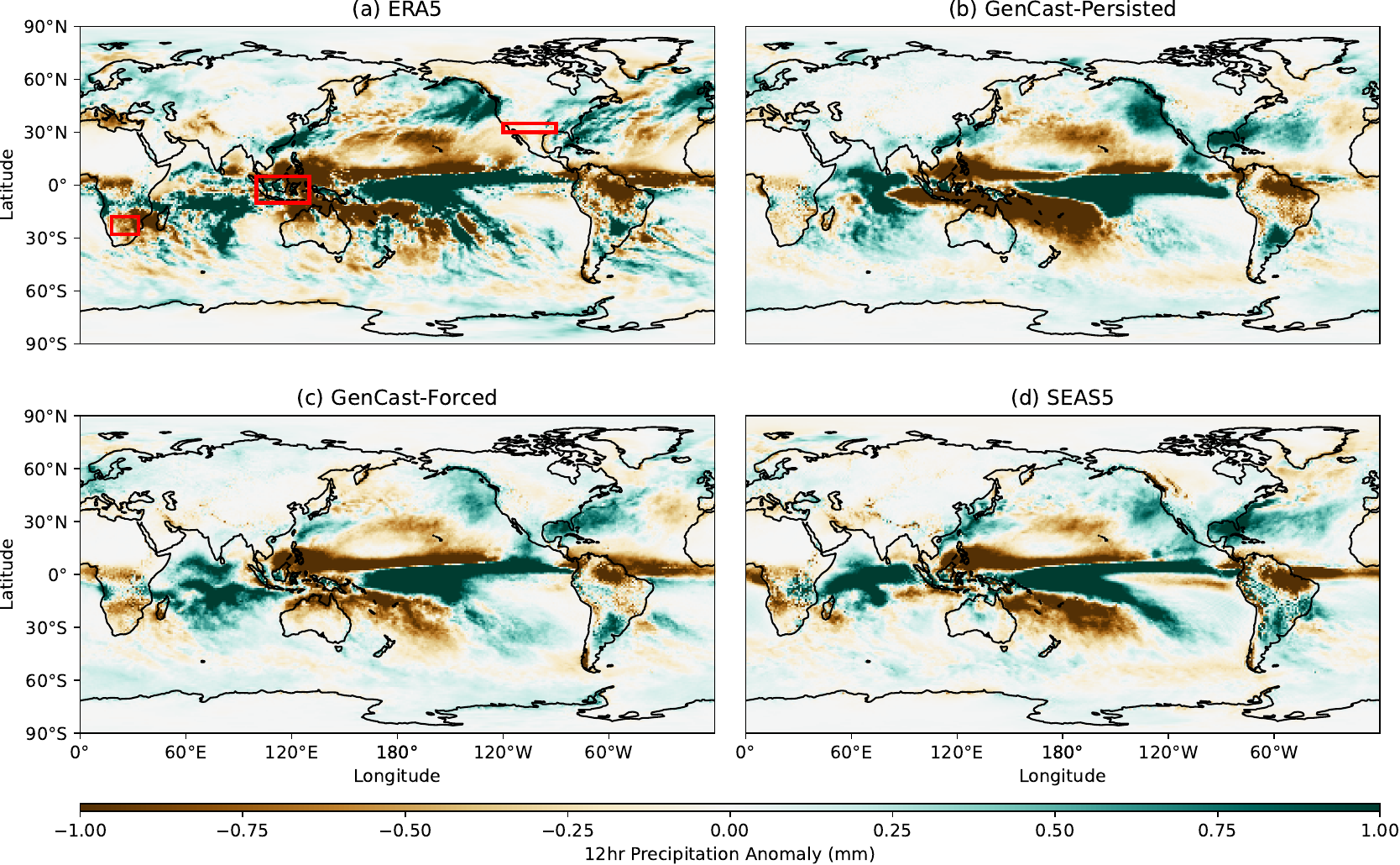}
       \caption{As for Fig.~\ref{fig:tp_anomalies_lanina}, but for December 2015 - February 2016 (Strong El Ni\~{n}o). }
    \label{fig:tp_anomalies_elnino}
\end{figure}

In order gain insight into how GenCast responds to sea surface temperature anomalies, we investigate precipitation forecasts produced when initialised in a year with strong El Ni\~{n}o or La Ni\~{n}a conditions. Since ENSO is one of the key ocean-related drivers of atmospheric variability \citep{mcphaden_enso_2006}, it is important that MLWP models can model the atmospheric response --- both mean and variability --- correctly. The periods chosen are December 2010 - February 2011 (strong La Ni\~{n}a conditions) and December 2015 - February 2016 (strong El Ni\~{n}o conditions), selected as they are the years with strongest ENSO signal. Whilst these periods are within GenCast's training period, this still represents an out-of-sample experiment since we are rolling GenCast autoregressively out to seasonal timescales far beyond the model's short training timescales. The resulting ensemble mean 12hr precipitation anomalies for December-February are shown in Figs.~\ref{fig:tp_anomalies_lanina} and \ref{fig:tp_anomalies_elnino}, using 20 ensemble members for both SEAS5 and GenCast.

For the 2010-2011 La-Ni\~{n}a forecast anomalies shown in Fig.~\ref{fig:tp_anomalies_lanina} (a), we can see that both GenCast-Persisted and GenCast-Forced produce a distinct pattern of drying over the tropical Pacific and wetting over the maritime continent, in agreement with ERA5 and SEAS5. Some difference can be seen between GenCast-Persisted and ERA5 in e.g.~the Hudson Bay and Indian Ocean, although this is rectified with GenCast-Forced. GenCast also captures the correct pattern of drying and moistening associated with La Ni\~{n}a away from the Pacific basin, for example over South America, and over central and southern Africa. Around the Gulf of Mexico, the drying signal is stronger in both GenCast experiments than in SEAS5.

For the 2015-2016 El Ni\~{n}o forecast anomalies in Fig.~\ref{fig:tp_anomalies_elnino} (b), we similarly see an agreement in the pattern of wetting and drying over the tropical Pacific and maritime continent for all models. GenCast-Persisted predicts erroneous wetting over the tropical Atlantic, which is much improved by forcing with ERA5 SSTs. Both GenCast models appear to show a more accurate representation of the wetting and drying pattern in the Northern Atlantic.

\begin{figure}[!ht]
    \includegraphics[width=\textwidth]{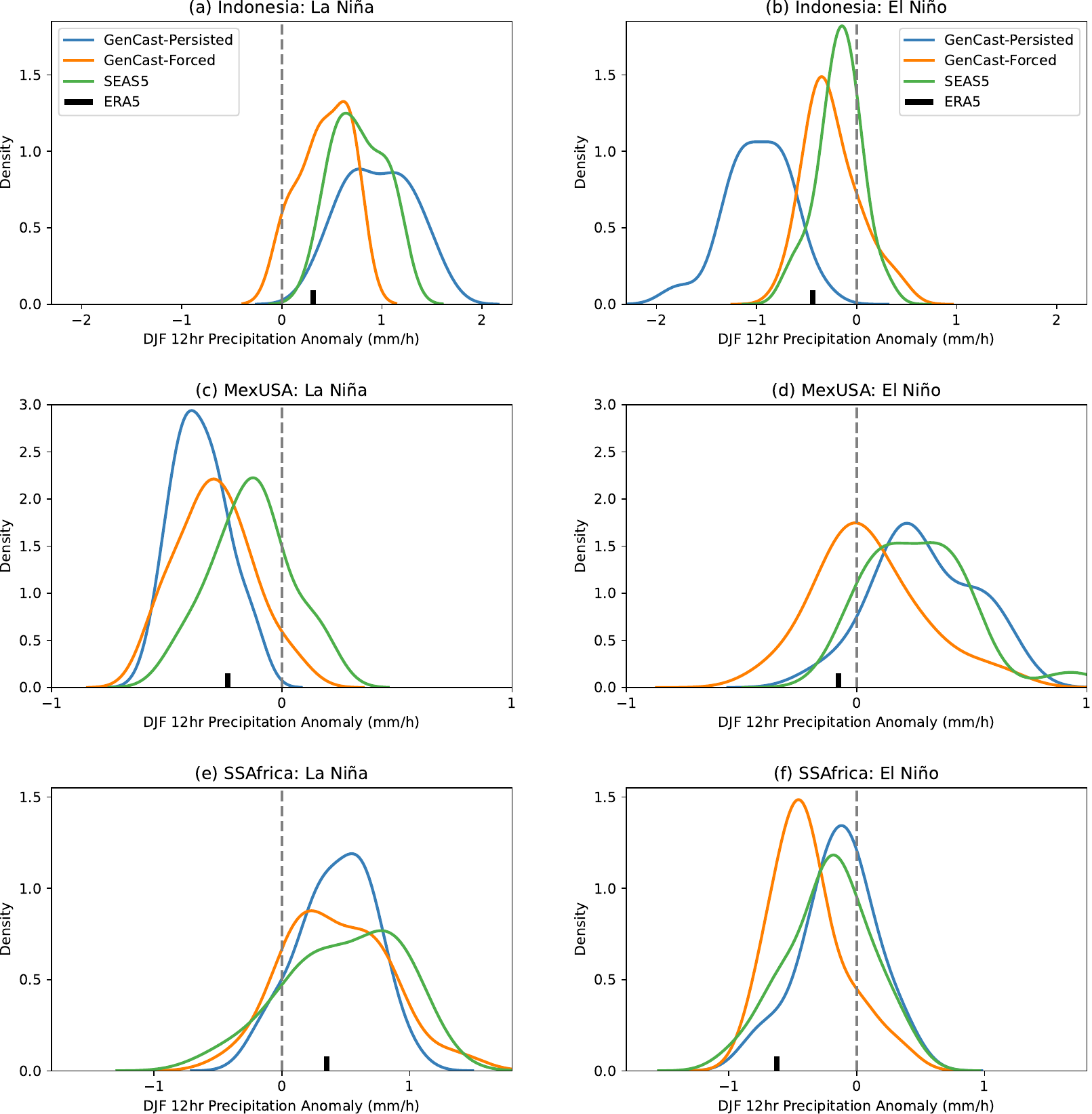}
       \caption{Distribution of DJF 12hr precipitation anomalies, averaged over each of the subregions shown in Figs.~\ref{fig:tp_anomalies_lanina} and \ref{fig:tp_anomalies_elnino} (rows), and for each of the La Ni\~{n}a and El Ni\~{n}o case studies (columns). In each plot the ERA5 value is shown in black.}
    \label{fig:tp_distribution}
\end{figure}

Apart from the ensemble mean prediction, it is also important to check that the ensemble distribution is reasonable compared to the physical forecast and observations. In Fig.~\ref{fig:tp_distribution} we show the distribution across the ensemble of 12 hour precipitation anomalies averaged over several regions, chosen from \cite{davey_probability_2014} as areas with a notable response to El Ni\~{n}o and La Ni\~{n}a in December-February (see Sec.~\ref{sec:data}). Overall we can see that GenCast-Persisted and GenCast-Forced produce distributions of a similar spread and mean value to SEAS5, with clear shifts across the 0 anomaly line between the La Ni\~{n}a and El Ni\~{n}o years, and such that the ERA5 data point (black line) lies within each of the distributions. For Indonesia in panels a and b (wetter/drier during La Ni\~{n}a/El Ni\~{n}o), we can see that the the bulk of the forecast distributions fall on the expected side, with GenCast-Persisted having the greatest variation between El Ni\~{n}o and La Ni\~{n}a, whilst the SEAS5 and GenCast-Forced distributions are fairly similar. For MexUSA in panels c and d (drier/wetter during La Ni\~{n}a/El Ni\~{n}o), all models show a shift to increased precipitation moving from La Ni\~{n}a to El Ni\~{n}o conditions, with GenCast-Persisted showing a particularly pronounced drying signal during the La Ni\~{n}a year. For SSAfrica in panels e and f (wetter/drier during La Ni\~{n}a/El Ni\~{n}o), all forecasts show clear drying and wetting signals, with similar shaped distributions, although GenCast-Forced showing a particularly pronounced drying signal during the El Ni\~{n}o year.

In summary, GenCast is able to capture the observed response to these two strong ENSO events, indicating that it has learned to correctly replicate some of the atmospheric response to sea surface temperature, despite this interaction not being a dominant driver in skill at the timescales it was trained at. This is also reflected in the change in distribution of precipitation averaged over several subregions; there is a clear shift in all of the forecast distributions that reflects the expected wetting and drying behaviour for each subregion in response to the ENSO conditions, and the distributions of both GenCast-Persisted and GenCast-Forced show similar spread and mean to the SEAS5 distribution.

\subsection{Anomaly correlation}
\label{sec:acc}
In this section we evaluate the skill of GenCast in predicting seasonal 2-metre temperature (2mT) and mean sea level pressure (MSLP), using the anomaly correlation coefficient (ACC).

\begin{figure}[!ht]
    \centering
        \includegraphics[width=\textwidth]{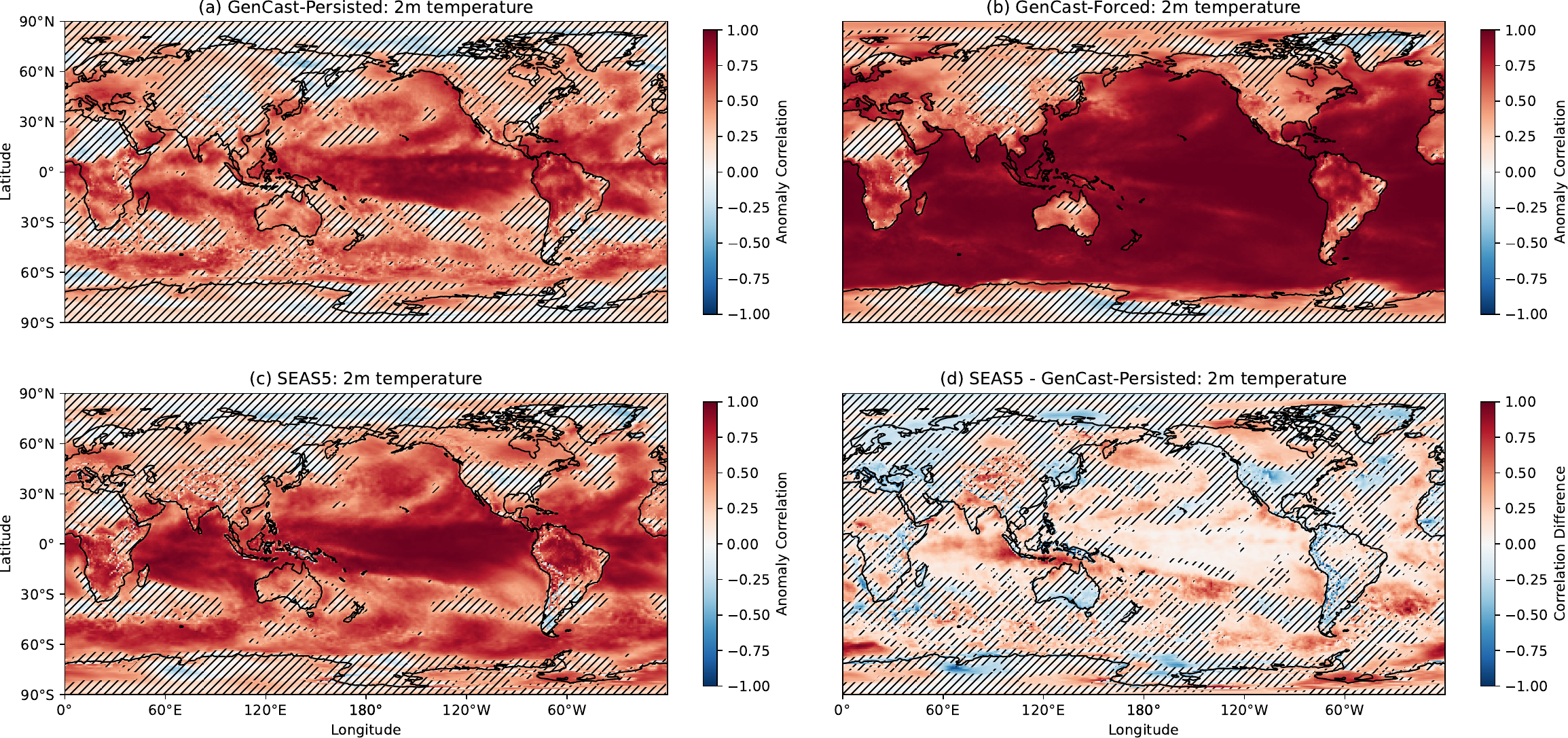}
       \caption{Anomaly correlation coefficient (ACC) for DJF 2-metre temperature, for forecasts initialised on 1\textsuperscript{st} November 2004-2024. Higher ACC indicates more skill. Hatching indicates where correlations or correlation differences are significant at the 95\% level (see Sec. \ref{sec:sig}).}
    \label{fig:2mt_acc}
\end{figure}

The ACC results for 2mT are shown in Fig.~\ref{fig:2mt_acc}. Panels a and c show that there are similar patterns of skill for GenCast-Persisted and SEAS5, with SEAS5 generally achieving higher correlation in the tropics and maritime continent. The high correlations achieved with GenCast-Forced over the ocean (panel b) show that GenCast is using the SST input appropriately to set the 2-metre temperature, whilst the low skill over much of the land points highlights the need for more land surface information in GenCast's inputs. Figure.~\ref{fig:2mt_acc} (d) confirms that there is no significant difference between the skill of SEAS5 and GenCast-Persisted over large parts of the Extra-Tropics, though SEAS5 is significantly more skilful over Tropical ocean regions. In contrast, GenCast-Persisted outperforms SEAS5 over some mountainous regions including the Andes, the Rockies and the Alps. It is also interesting to note that GenCast-Persisted shows a slight improvement in the North-West Atlantic, a feature attributed to how SEAS5 captures the variability of the North Atlantic subpolar gyre \citep{johnson_seas5_2019}.

Since GenCast does not does not receive any information about the sea ice, such as ice concentration or temperature, we might expect significant differences over areas of high sea ice concentration. Whilst SEAS5 does seem to perform significantly better over over some of the Wedell and Beaufort seas, there are also some areas, such as around the Anzhu islands and in the Ross sea, for which GenCast-Persisted achieves higher skill. The ACC results for 2mT aggregated by region are shown in Table \ref{tab:2mt}. From this we can see that overall GenCast-Persisted performs comparably to SEAS5 in the Northern extratropics, with significant differences in the tropics and Southern extratropics.

\begin{figure}[!ht]
    \includegraphics[width=\textwidth]{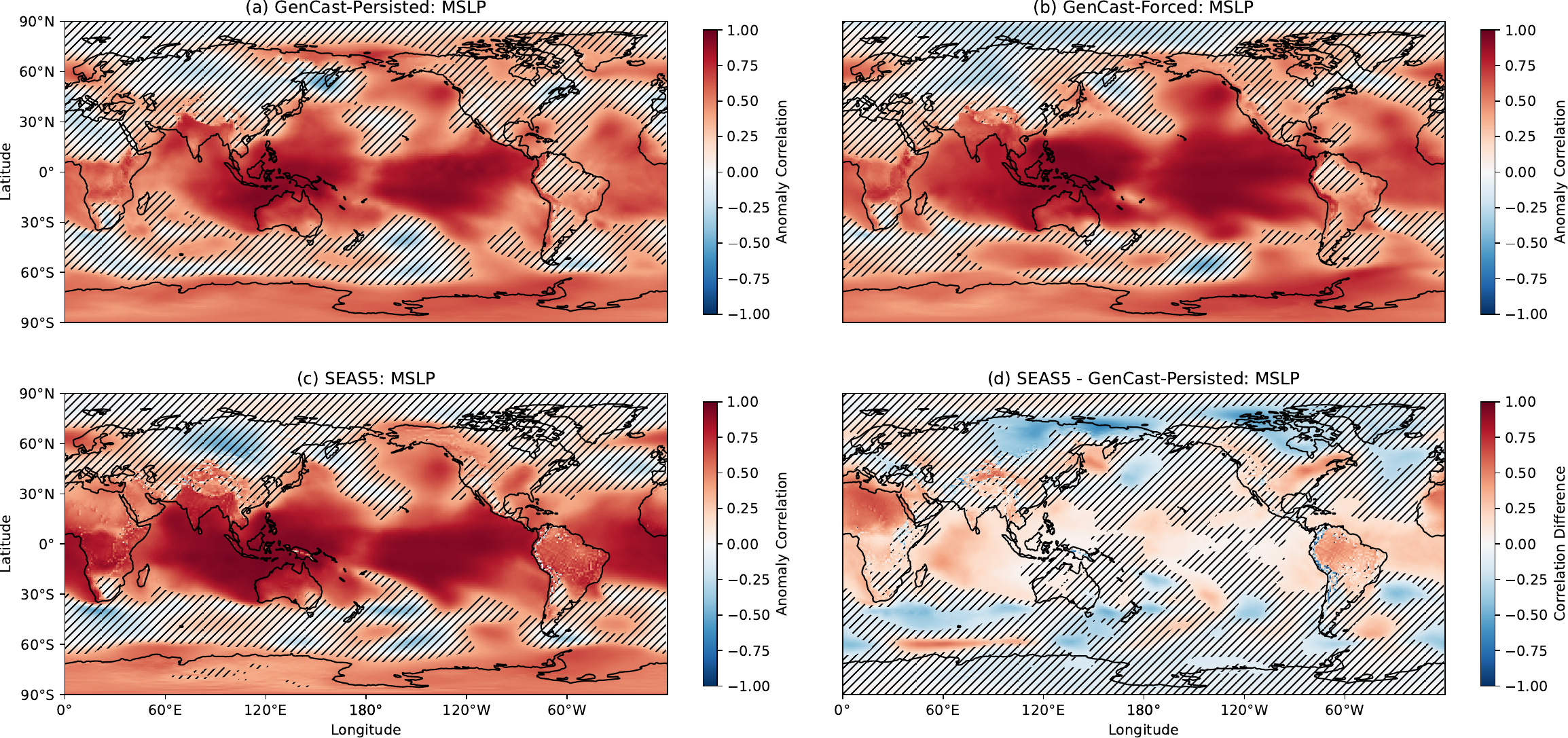}
       \caption{Anomaly correlation coefficient (ACC) for DJF mean sea level pressure, for forecasts initialised on 1\textsuperscript{st} November 2004-2024. Higher ACC indicates more skill. Hatching indicates where correlations or correlation differences are significant at the 95\% level (see Sec. \ref{sec:sig}).}
    \label{fig:mslp_acc}
\end{figure}

The ACC results for MSLP are shown in Fig.~\ref{fig:mslp_acc}. Again all forecasts show similar patterns of skill, with SEAS5 significantly outperforming GenCast-Persisted over northern Africa, South America, the Sea of Okhotsk, and the tropics, as shown in panel d. There are some areas in the midlatitudes where GenCast-Persisted appears to improve upon SEAS5, such as over northern Canada, northern Asia and the North Atlantic ocean. There is a less pronounced difference between GenCast-Persisted and GenCast-Forced, suggesting that an accurate representation of the ocean is not sufficient to achieve much higher skill for this field, or perhaps reflecting the importance of atmosphere-ocean coupling in these regions. The ACC results for MSLP aggregated by region are shown in Table \ref{tab:mslp}. From this we can see that differences in skill between GenCast-Persisted and SEAS5 are concentrated in the tropics, with the two models performing similarly in the extratropics.

\begin{table}[ht]
\centering
\caption{Anomaly correlation coefficient results for 2mT aggregated by region.}
\begin{tabular}{c  c  c  c } 
 \hline
  & \multicolumn{1}{p{2.5cm}}{ \centering 2mT \\  Tropics  } & \multicolumn{1}{p{2.5cm}}{ \centering 2mT Northern\\  Extratropics  } & \multicolumn{1}{p{2.5cm}}{ \centering 2mT Southern\\  Extratropics  } \\ 
 \hline\hline
GenCast-Persisted & 0.62  & 0.25 &  0.42 \\ 
GenCast-Forced & 0.88  & 0.51  & 0.76 \\    
SEAS5 & 0.74  & 0.28 & 0.54 \\   
 \hline
\end{tabular}
\label{tab:2mt}
\end{table}

\begin{table}[ht]
\centering
\caption{Anomaly correlation coefficient results for MSLP aggregated by region.}
\begin{tabular}{c  c  c  c } 
 \hline
  & \multicolumn{1}{p{2.5cm}}{ \centering MSLP \\  Tropics  } & \multicolumn{1}{p{2.5cm}}{ \centering MSLP Northern\\  Extratropics  } & \multicolumn{1}{p{2.5cm}}{ \centering MSLP Southern\\  Extratropics  } \\ 
 \hline\hline
GenCast-Persisted & 0.62  & 0.22  & 0.48  \\ 
GenCast-Forced & 0.71  & 0.21 & 0.54 \\    
SEAS5 & 0.76  & 0.21 & 0.50 \\   
 \hline
\end{tabular}
\label{tab:mslp}
\end{table}

\subsection{NAO prediction}

\begin{figure}[!ht]
    \centering
    \includegraphics[width=\textwidth]{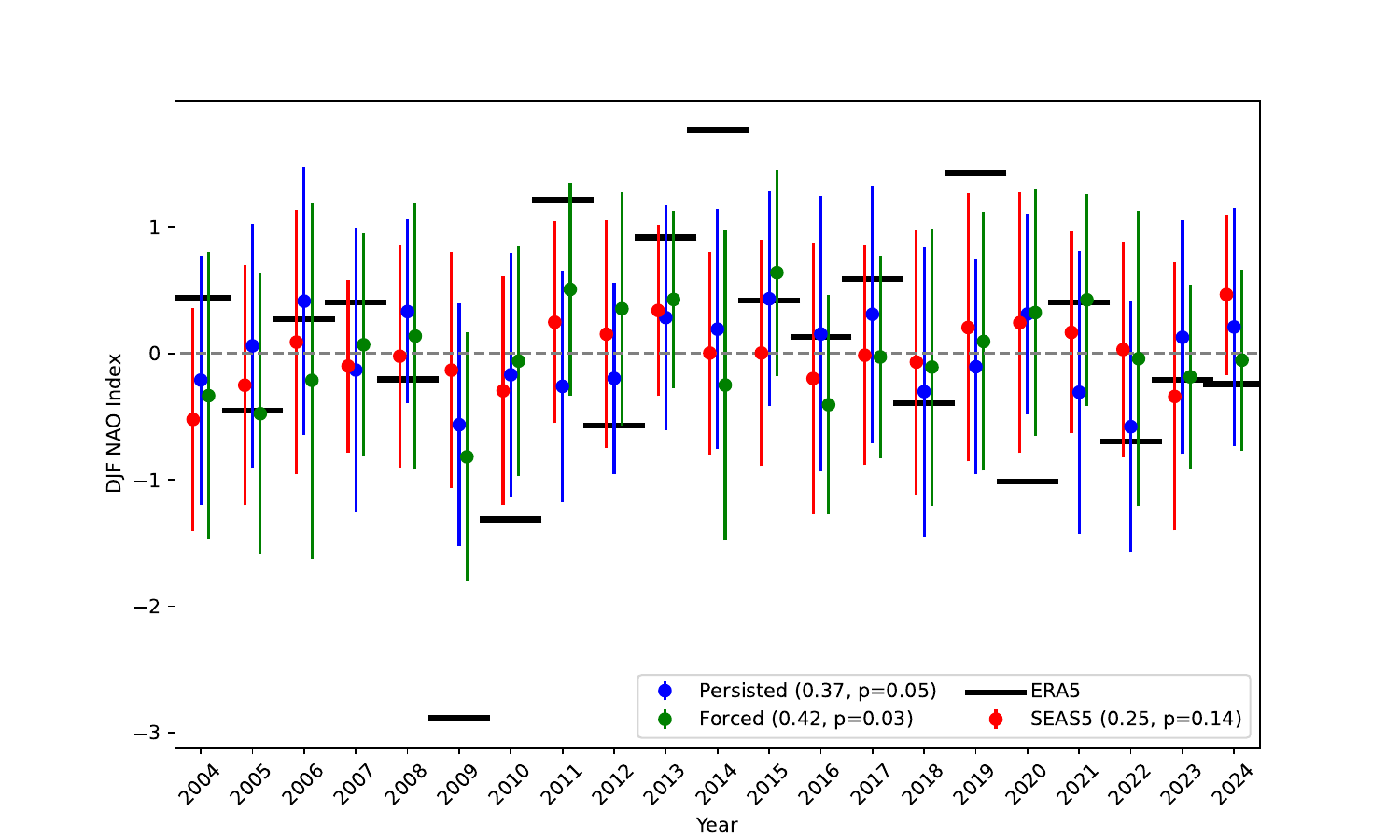}
    \caption{NAO Index calculated from the mean sea level pressure predictions of GenCast, aggregated by season. Error bars indicate the spread of the ensemble members. Numbers in brackets are the Pearson correlation of each time series with ERA5, including an estimated p-value.}
    \label{fig:nao}
\end{figure}

In this section we evaluate how well GenCast predicts the North Atlantic Oscillation (NAO), which is an important driver of weather and climate variability in Eurasia and North America \citep{hurrell_overview_2003}. The predictions of the NAO index are shown in Fig.~\ref{fig:nao}, where each time series has been centred by subtracting its mean over the 20 year period, and normalised by dividing by the standard deviation of the index calculated on ERA5 data.

SEAS5 systematically underestimates the variability of NAO values compared to observations, with a correlation of just 0.25 (not significant at the 5\% level). This is related to the so-called `signal-to-noise' problem, a problem shared by all physical models capable of performing skillful NAO forecasts \citep{scaife2018signal, johnson_seas5_2019}. Interestingly, the MLWP forecasts share the same issue, consistent with the result of \cite{watt-meyer_ace2_2025}. With regards to the skill, it is noteworthy that GenCast-Persisted obtains a higher correlation of 0.37 (significant at the 5\% level) with ERA5 than SEAS5 does over this time period. There is also only a small improvement realised with GenCast-Forced (significant correlation of 0.42), consistent with the relatively small difference seen for the MSLP ACC results in Sec.~\ref{sec:acc}.

\subsection{Reliability of the ensemble}

\begin{figure}[!ht]
    \centering
    \includegraphics[width=\textwidth]{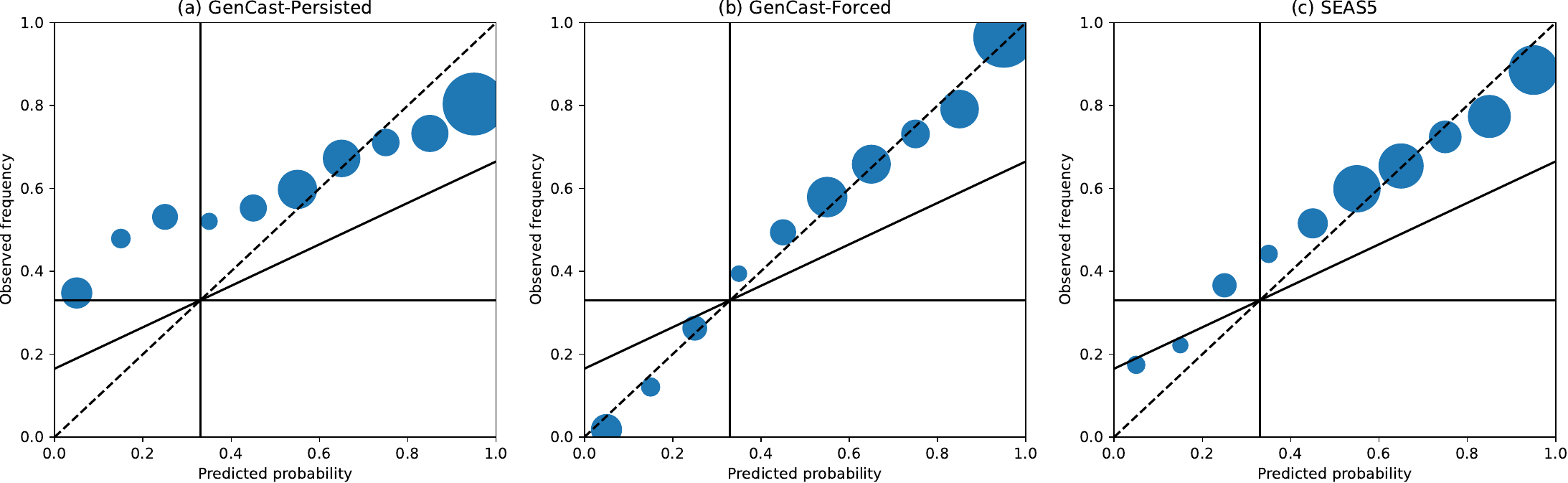}
       \caption{Reliability for each model, where the forecast objective is to predict the probability of 2-metre temperature in December-February being above the lower tercile. (a) GenCast-Persisted, (b) GenCast-Forced, (c) SEAS5. The black dashed line in each plot indicates the line a perfectly reliable forecast would lie along. Circle sizes indicate the number of examples within each probability bin, and circles are shown at the centres of each probability bin. }
    \label{fig:reliability}
\end{figure}

For any probabilistic forecasting system it is important that the forecast probabilities are good indicators of how likely an outcome is, in order for the forecast to have value to decision makers . Whilst the skill of a probabilistic forecast can be summarised by one of many forecast skill metrics, a reliability diagram provides a fuller insight into the joint probability distribution of the forecast and observations for a particular binary event of interest \citep{wilks_statistical_2011}. Reliability diagrams for the seasonal forecasts are shown in Fig.~\ref{fig:reliability}, where we compare the forecast probability and observed frequencies of the seasonal 2m-temperature being above the lower tercile. 

The reliability diagram of GenCast-Persisted, in  Fig.~\ref{fig:reliability} (a), shows that it is overconfident in its predictions, and particularly deviates from the optimal dashed line at low predicted probabilities. SEAS5, shown in panel (c), shows a significant improvement, particularly for points with low predicted probability. For GenCast-Forced we can see that the reliability is very well aligned with the dashed line, more so than SEAS5; this indicates that GenCast combined with a realistic representation of the ocean variability may be enough to produce a well-calibrated probability distribution of these events.

\section{Discussion}

In this work we have demonstrated the first application of GenCast to seasonal forecasting, far beyond the timescale it was trained at, by running the model for 4 months with prescribed sea surface temperature (SST) boundary conditions: In the first setup, GenCast receives persisted SST anomalies (GenCast-Persisted), whilst in the second setup GenCast receives SSTs from ERA5 (GenCast-Forced). Whilst these are not full seasonal forecasts, as they lack an interactive ocean, they provide a test into how well GenCast has learned to model long term physical processes having being trained on single timestep predictions and optimised for the medium-range. The model produces a 4-month forecast in around half an hour on a single A100 GPU, compared to around 3 hours on 50 cores for the IFS at a similar resolution \citep{mogensen_effects_2018}; whilst it is not as fast as some deterministic MLWP models, it still offers the potential to achieve higher ensembles much more efficiently than SEAS5.

An evaluation of precipitation for two years with strong El Ni\~{n}o / La Ni\~{n}o SST warming patterns show that GenCast is able to capture the systematic patterns of wetting and drying appropriately, in some areas perhaps more accurately than SEAS5. An investigation of the distribution of 12hr DJF precipitation anomalies averaged over three particular subregions also demonstrated distinct shifts in distribution in response to the ENSO conditions, with distributions that aligned well with SEAS5 in terms of mean value and spread. 

Anomaly correlations of 2-metre temperature calculated over the full 20-year dataset reveal that, whilst SEAS5 tends to achieve higher skill in many areas, particularly in the tropics, several areas in the extratropics and some mountainous regions appear to exhibit some improvement in skill. Whilst there are areas of high sea ice concentration, such as the Weddell sea, for which SEAS5 achieves significantly higher skill, GenCast-Persisted achieves high skill in some regions such as the Ross sea, which is perhaps surprising since it receives no information about the sea ice concentration or temperature. GenCast-Forced achieves very high correlation over the ocean points, which confirms that SST input is being used appropriately to inform the 2-metre temperature. Over land points, however, there are still many areas where GenCast-Forced achieves low correlation, highlighting the need for more land surface information to be included in the model inputs.

Anomaly correlations of mean sea level pressure (MSLP) show that SEAS5 achieves superior skill in the tropics, although in the midlatitudes there are areas such as Siberia and northern America for which GenCast-Persisted achieves higher skill. This is reflected in forecasts of the North Atlantic Oscillation (NAO) index, for which GenCast-Persisted achieves higher correlation with the NAO index calculated using ERA5. Differences in skill between GenCast-Persisted and GenCast-Forced for MSLP are relatively small, indicating that accurate ocean information alone is not enough to drive skill in this model. Instead, a coupled ocean or additional variables may be needed. We note that GenCast-Persisted has a lower correlation than that reported over 1994-2016 using ACE2 \citep{kent_skilful_2025}. Unlike GenCast, ACE2 receives information about sea ice as an input, which could be a source of the skill difference between the two models.

Finally we investigate the reliability of probabilistic predictions of 2-metre temperature being within the lower tercile. GenCast-Persisted shows overconfidence in its probabilities compared to SEAS5, particularly for low probability events. However, the probabilities for GenCast-Forced are very well calibrated, indicating that the missing variability in the sea surface temperature may be enough to produce well calibrated ensembles with GenCast.

We acknowledge several limitations of this study. Firstly, it is common with seasonal forecasts to perform hindcasts in order to correct biases and drifts in the forecasts. Since we have not performed this step for either the GenCast or SEAS5 forecasts, we cannot say to what extent differences in performance are related to different lead-time dependent biases, or which forecast would perform better if such biases were removed. 

Since the sea surface temperatures are prescribed, this is also not a demonstration of full seasonal forecasting, but an indication of how well GenCast has learned to model long-term dynamics having been trained at short timescales. In future work we intend to explore how coupling to a full dynamic or machine-learnt ocean model will change GenCast's seasonal forecasting skill and reliability.

Overall, the results show promise in the use of generative models such as GenCast to perform seasonal forecasts, providing further validation as to how well the model has learnt to capture physical processes. It can reproduce the atmospheric response to drivers of variability on seasonal timescales, despite the limited role of these drivers on variability at the training timescale of 12 hours. The results motivate the further study of models similar to GenCast coupled with a dynamical or machine-learnt ocean model. 

\bmhead{Acknowledgements}

This publication is part of the EERIE project funded by the European Union (Grant Agreement No 101081383). Views and opinions expressed are however those of the author(s) only and do not necessarily reflect those of the  European Union or the European Climate Infrastructure and Environment Executive Agency (CINEA). Neither the European Union nor the granting authority can be held responsible for them. This work was funded by UK Research and Innovation (UKRI) under the UK government’s Horizon Europe funding guarantee (grant number 10049639).  Acknowledgement is made for the use of ECMWF's computing and archive facilities in this research. HMC was also funded through a Leverhulme Trust Research Leadership Award.

\bibliography{references}

\end{document}